# One-Way Quantum Key Distribution System based on Planar Lightwave Circuits

Yoshihiro NAMBU, Ken'ichiro YOSHINO and Akihisa TOMITA[1]

Fundamental and Environmental Research Laboratories, NEC Corp., 34 Miyukigaoka, Tsukuba, Ibaraki 305-8501, Japan

[1] ERATO-SORST Quantum Computation and Information Project, JST, Daini Hongo White Bldg. 201, 5-28-3 Hongo, Bunkyo, Tokyo 133-0033, Japan

We developed a one-way quantum key distribution (QKD) system based upon a planar lightwave circuit (PLC) interferometer. This interferometer is expected to be free from the backscattering inherent in commercially available two-way QKD systems and phase drift without active compensation. A key distribution experiment with spools of standard telecom fiber showed that the bit error rate was as low as 6% for a 100-km key distribution using an attenuated laser pulse with a mean photon number of 0.1 and was determined solely by the detector noise. This clearly demonstrates the advantages of our PLC-based one-way QKD system over two-way QKD systems for long distance key distribution.
KEYWORDS: quantum cryptography, integrated optics, optoelectronics

# 1. Introduction

Quantum key distribution (QKD) provides a way for two distant parties, say, Alice and Bob, to share a secret key using a communication channel [1-3]. The maximum information leaked to an eavesdropper, say, Eve, can be reliably estimated from the errors introduced in the distributed raw keys on the basis of quantum mechanics, such as the complementarity of two conjugate variables and the no-cloning theorem [4]. Introduced errors and leaked information in the distributed raw keys can be eliminated using classical post-processing including error correction and privacy amplification [5-6]. The resulting shared final keys have negligible leaked information and can be used for one-time pad cryptography.

After Bennett and Brassard invented [1] and demonstrated [2] the best-known QKD protocol, BB84, numerous QKD systems were developed using optical-fiber techniques and faint laser pulses [7-23]. In these systems, two weak coherent pulses (WCP) with a mean photon number of much less than one and with some time delay are used to carry the key data. The information is encoded on the relative phase delay between these pulses. Townsend et al. first demonstrated a QKD prototype using double asymmetric Mach-Zehnder interferometers (AMZIs) connected serially with optical fiber and 10-km transmission of single interfering photons [7]. Although this system is simple enough, it is difficult to apply to a practical QKD system because of the instability due to phase drift in the two independent and remote AMZIs. This instability can be eliminated using an active compensation technique, but the resulting system is complicated and expensive [19].

This difficulty was solved by Muller et al. [11] They used a single asymmetric Michelson interferometer and Faraday mirrors to develop an interferometer in which optical pulses make a two-way trip between two users. This two-way interferometer makes it possible to auto-compensate both phase and polarization drift. It is practical because it uses only existing technologies. Recently, several commercial prototypes based on this system have been released [24]. Despite its usefulness, several drawbacks originating from two-way photon transmission have been pointed out [14,17,19,21,23,25]. For example, Ribordy et al. noted that Rayleigh backscattering caused by the refractive index inhomogeneities of the link fiber limited the performance of their two-way QKD system (called a plug-and-play system) [14]. In the two-way system, if pulses traveling to and from Alice collide in the line, backscattered photons from an intense pulse propagating to Alice can accompany a weak pulse propagating back to Bob and increase the error rate considerably. This limits the transmission distance as well as the system repetition rate, i.e., the key rate, for two-way QKD systems. In fact, Subacius et al. extensively investigated the backscattering in the fiber and its impact on the performance of a two-way QKD system. They concluded that for the two-way QKD system, a longer transmission and higher key rate are conflicting demands, even if the photon detector performance is improved [25]. Alternatively, the collision of crossing pulses in the fiber can be reduced by sending bursts of pulses spaced by long dead intervals and using a storage line in Alice's system, but this eventually reduces the effective key rate [14]. Several researchers have also noted that a two-way system is likely to be more vulnerable to a Trojan horse attack on Alice's apparatus and incompatible with highly secure systems using true single photon sources [19,21].

An alternative approach to the aforementioned problems based on a one-way interferometer using integrated-optic AMZIs [26-29] was recently reported. By applying a planar lightwave circuit (PLC), the instability due to the phase drift inherent in a one-way QKD system and the polarization drift in the link fiber were eliminated without introducing any complex active compensation mechanisms. We showed that this system is sufficiently stable and has the potential to enlarge the key distribution distance by demonstrating single-photon interference over 150 km [29]. In this paper, we report on the QKD system using this PLC-based interferometer and on our demonstration of key distribution over distances up to 100 km in the laboratory. Our experimental results show that the quantum bit error rate (QBER) of our system was small enough for key distillation and depended solely on the detector noise. This implies that our system is free of backscattering and is stable with no active compensation, demonstrating its practical importance for long distance key distribution.

## 2. PLC-Based One-Way QKD System

A schematic of our one-way QKD system is shown in Fig. 1. The system is based on a time-division interferometer consisting of double AMZIs [7]. The details of the interferometer are in ref. [29]. The AMZIs, which are $4.9 \times 8.9$ cm$^2$, were fabricated on a silica-based PLC platform [30]. They each have a 5-ns delay in one of their arms and an excess optical loss of <1.5 dB. The branching ratio of one of the couplers was made asymmetric to compensate for the difference in the optical loss between the two arms. A Peltier cooler attached to the back of the substrate controlled the temperature of the device with a precision of up to 0.01°C. As shown later, this interferometer can be made insensitive to polarization drift in the link fiber by appropriately choosing the temperature of Bob's AMZI.

To follow QKD protocol, we introduced two phase modulators in Alice's station and one in Bob's. By applying pulsed modulations as described in the following, the quantum communication required for four-state protocols such as BB84 and the recent SARG04 is enabled [31,32]. Short pulses with pulse widths of ≤200 ps from a 1.55-μm distributed feedback laser diode (LD) were fed into the encoding AMZI. The LD was synchronized with Alice's master controller (MC) and operated at 1 MHz, which is low enough to avoid after-pulsing noise in the APD photon detector [33]. The relative phase between coherent double pulses emerging from the AMZI was modulated with a series of inline phase modulators in Alice's apparatus. Alice selects two independent random bit strings generated and stored in her MC. At the first phase modulator (PMA$_1$), a 0 or $\pi$ phase shift is applied to one of the double pulses according to her first random bit, while at the second modulator (PMA$_2$), 0 or $\pi/2$ is applied according to her second random bit. To make these modulations, a wideband LiNbO$_3$-phase modulator (Sumitomo Osaka Cement: T.PM 1.5-2.5-P-FK) driven by short electric pulses (lasting approximately 1 ns) was used. The polarization of the input pulses and optic axes of all the devices in Alice's apparatus are matched so that the polarizations of double pulses emerging from the PMA$_2$ are the same and are linearly polarized. Then, the pulses are strongly attenuated so that, on average, 0.1 photons per clock cycle leave Alice's apparatus.

The pulses propagate along the fiber, and the polarization state changes. However, double pulses experience the same change and thus have the same polarization state at the fiber output as long as the time interval of the double pulses is much shorter than the time scale of the polarization drift. This condition is satisfied because changes in temperature and/or mechanical stress have larger time constants than the pulse interval. Then, double pulses with the same polarization enter Bob's apparatus. Bob selects a random bit string generated and stored in his slave controller (SC). According to his random bit, a 0 or $\pi/2$ phase shift is applied to one of the double pulses at the third phase modulator (PMB$_3$). After traveling through a decoding AMZI, photons can be found in three timeslots in each of two output ports. Half of the photons, which are found in the first and last timeslots, are independent of the relative phase between the propagating double pulses and discarded. The other half, found in the middle timeslot, depends on the relative phase and contributes to key generation. Photons found in this timeslot are selectively detected by balanced, gated-mode InGaAs/InP avalanche photodiodes (APDs) operating at approximately −100°C [34] and registered by Bob's SC. This SC and the gated photon detectors must be synchronized with Alice's MC. For this reason, we used the fiber optic ribbon used in optical fiber cables to link Alice and Bob. One core (the transmission fiber) was used to transmit QKD signals; the other one (the sync fiber) carried strong laser pulses used to synchronize Alice and Bob. Both the QKD signals and the synchronization pulses were initiated by the same source LD.

3. **Basic Performance of PLC-Based Interferometer**

Nearly perfect classical interference was observed for our PLC-based interferometer, as shown in the inset in Fig. 2. To observe the interference, two AMZIs were connected using a short single-mode fiber, and relatively intense pulses were introduced. The photon count rates in the two output ports of the decoding AMZI were observed while controlling the relative phase delay of the encoding AMZI by controlling its temperature ($T_A$). To demonstrate the stability of our PLC-based interferometer, we kept both AMZIs at the temperature at which maximum interference was observed initially (shown by a double-pointed arrow in the inset) and recorded the count rates in the two ports over 3 h. As shown in Fig. 2, our interferometer has visibility and stability enough to distribute quantum keys without complicated active compensation, demonstrating that it is robust against the phase drift. The slight decrease in the measured ratio might be due to the quite slow and independent temperature drift in the two AMZIs. In addition, our interferometer can be made insensitive to polarization drift in the optical link [29]. Because double pulses with the same polarization enter Bob's apparatus, balancing the birefringence of the two arms in Bob's AMZI suffices for polarization insensitivity. This can easily be achieved by controlling the temperature of the decoding AMZI ($T_B$) as shown in Fig. 3. The optimal temperature, around 18.3°C, was found from the fringe visibility measured by changing $T_A$. At this temperature, the difference between the modal phase shift in the long (*L*) and short (*S*) arms of the AMZI ($\theta_L - \theta_S$) is a multiple of $2\pi$. We found that the fringe visibility remained at this temperature even if we intentionally disturbed the polarization of the optical link. This birefringence balancing was much less sensitive to the device temperature than the interference was. Therefore, we could easily simultaneously balance the phase setting and the birefringence. In the QKD experiment described below, we achieved highly stable key generation by controlling the temperature of two AMZIs independently around 18.3°C with 0.01°C precision.

To make the QKD system insensitive to the polarization drift in the optical link, we used a polarization-insensitive phase modulator for $PMB_1$ in Bob's site that is based on the polarization diversity technique [35,36]. We constructed a looped polarization interferometer in which orthogonally polarized light counterpropagates through polarization-maintaining fiber and put a low-loss $LiNbO_3$-phase modulator (EOSPACE) in it. The price paid was increased loss, which was approximately 3.2 dB.

## 4. Key Distribution Experiment

To evaluate the performance of our system, we distributed a key over 100 km using the BB84 protocol. Sufficiently large numbers of weak pulses were transmitted from Alice to Bob, each of which was in one of the BB84 states. After this transmission, Alice and Bob made a basis reconciliation to find the pulses for which their selected basis agreed. This was carried out by revealing Alice's second random bit used for $PMA_2$ and Bob's random bit used for $PMB_1$. From Alice's first random bit used for $PMA_1$ and the bit assigned to the output port in which Bob found a photon, Alice and Bob extracted a subensemble of data associated with the photons found. These data should be perfectly correlated if no eavesdropper is present and if the system has no imperfections and thus constitute a shared random key. Eve cannot get information on the key without introducing errors in the correlations between Alice and Bob.

Fiber transmission experiments were undertaken using spools of fiber optic ribbon, with Alice and Bob located in the same room. The results are shown in Fig. 4. The top graph shows the key generation rate for several fiber lengths. The solid symbols indicate the sifted bit rate before classical post-processing, while the open symbols indicate the final bit rate after the classical post-processing described in the following. The circle and square symbols indicate data associated with different conditions of the photon detector, that is, quantum efficiencies (QE) of 5 and 10%. The sifted bit rate for a fiber length of 0 km was determined by multiplying the system repetition (1 MHz), mean photon number per clock cycle (0.1), QE of the photon detector, loss in Alice's apparatus, and a systematic factor of 1/4 due to noninterfering photon contribution and yield of sifting. The sifted bit rate decreased with increasing fiber length at a rate of approximately 0.205 dB/km, which is slightly larger than the measured value of the fiber used of approximately 0.2 dB/km. This is probably due to pulse width broadening caused by dispersion in the fiber and the finite opening window of the APD gating (approximately 1.2 ns).

The sifted key involved errors, which might have been caused by eavesdropping as well as imperfections in the system. The QBER was measured by directly comparing portions of the yielded sifted keys of Alice and Bob with lengths of at least 4000 bits. The QBER is defined as the ratio of the number of erroneous bits in the sifted key to the total number of sifted bits. When the fiber is short, the measured QBER, shown in the bottom graph in Fig. 4, is independent of the QE of the photon detector and is around 1%. The measured QBER increases with increasing fiber length and depends considerably on the QE. It was as low as 6% for a 100-km fiber with QE=5%. This result can essentially be explained using the following simplified model. We consider only two major sources of errors. The first is the detector noise, which is characterized by the dark count probability per gate pulse, $d$, of each detector. Let $R$ be the sifted key generation rate and $D$ be the rate at which erroneous bits are generated; this means QBER=$D/R$. Then, the contribution of the dark count to $D$ is $2 \times 1/2 \times 1/2 \times d = d/2$, where the factor of 2 accounts for the number of detectors and the two factors of 1/2 are related to the fact that a dark count has a 50% chance of occurring when Alice and Bob have chosen incompatible bases (which are eliminated during sifting) and a 50% chance of occurring in the correct detector [3]. On the other hand, the contribution to $R$ is $2 \times 1/2 \times d = d$ because the last factor of 1/2 must be omitted for $R$. The second source of error is the optical imperfection of the interferometer, which is characterized by the classical interference visibility, $V$, or the extinction ratio, $e$, of the interferometer. Let us consider the QBER due to the optical imperfection and designate it QBER$_{opt}$, i.e., QBER$_{opt}$=$e/(1+e)$=$(1-V)/2$ [3]. Then, the contribution of the optical imperfection to $D$ is QBER$_{opt}(R-d)$. Finally, we have

$$\text{QBER} = \frac{\text{QBER}_{opt}(R-d) + d/2}{R} = \text{QBER}_{opt} + \frac{1-e}{1+e}\frac{d}{R}, \qquad (1)$$

where the first term represents the contribution of the optical imperfection, and the second one, the contribution of the detector noise. To account for the experimental results in Fig. 4, we need to clarify the dark count probability of our detector. The measured value of $d$ as a function of the QE of our detector is shown in Fig. 5.. The detector has characteristics similar to reported ones; i.e., the data can be fit well by an exponential curve [33]. Because of this nonlinear relationship between $d$ and QE, we can increase the signal to noise (SN) ratio of our detector at the expense of QE, which reduces the errors caused by the dark count. The broken lines in the bottom graph of Fig. 4 show the QBER calculated from eq. (1) using the measured value of $d$ and assuming that $QBER_{opt}$ is 1%, which corresponds to the extinction ratio of 20 dB. The agreement between the experiments and the calculations is almost perfect for the two experiments, although the assumed $QBER_{opt}$ is slightly larger than the actual value (see Fig. 2). This is due to the contribution of after-pulsing noise in the APD, which was observed in some systems [19]. Actually, we have found from a separate experiment that the probability of after-pulsing was about 5% for our APD operating at approximately −100°C and with 1 MHz repetition. Taking into account the fact that a factor of 1/4 of the after-pulsing noise contributes to the QBER (a factor of 1/2 contributes to the sifted keys in which a factor of 1/2 results in errors), the observed QBER for a short transmission length is consistent with this experimental result. From the above discussion, we conclude that the dominant contribution to the QBER originates from the after-pulsing noise of the APD for short fibers, whereas it originates from the detector noise for long fibers. Thus, this result demonstrates that our one-way QKD system is free from backscattering, as predicted.

The final key generation rates can be estimated using the sifted key generation rates and the QBER obtained in the experiment. To obtain a final key, we first applied a cascade error correction routine [37], followed by privacy amplification [38] using a Toeplitz matrix [39] to eliminate the information on the shared key gained by Eve. The privacy amplification requires a model that relates the QBER and estimates of the upper limit of information leaked to Eve. Here, we used a model for single-photon QKD, which has been established theoretically [40]. The broken lines in the top graph of Fig. 4 show the estimated final key generation rates. The figure indicates that the upper limit of the key distribution distance is 98 km for QE=10% and 118 km for QE=5%. These upper limits agree approximately with those estimated from the security limit for a QBER of less than 11.5%, which was established for single-photon QKD [41]. Thanks to an improved SN ratio, the upper limit of the key distribution distance was higher for QE=5% than QE=10% at the cost of reducing the key generation rate. This upper limit depends on the performance of the photon detector. It can be further increased if the SN ratio of the detector is further improved. For example, if we can reduce the dark count probability by one order of magnitude while keeping QE=10%, the upper limit will reach 147 km.

Although the QKD scheme shown above may provide us with a reasonable level of security against eavesdroppers restricted by the current technologies, it is not unconditionally secure. Unconditional security is possible if we use a single photon source, but QKD using WCP is well-known to be vulnerable to the photon-number splitting (PNS) attack [42,43]. Using this attack, Eve gets a large amount of information about the bits without introducing errors. Although the possible distance is limited considerably, unconditional security can be recovered if we carefully control the mean photon number according to the transmission distance [21,22,43,44]. It is a matter of course that our system is compatible with the SARG04 protocol that uses the same BB84 states but is more robust against PNS attacks [31,32]. To follow this protocol, we only need to exchange the roles that the three random bits and the bit assigned to the output port of Bob's AMZI play in the BB84 protocol; i.e., Alice's first random bit used for $PMA_1$ and the bit assigned to the output port in which Bob found a photon should be revealed during the reconciliation, while Alice's second random bit used for $PMA_2$ and Bob's random bit used for $PMB_1$ yield the key data. By carefully controlling the mean photon number according to the transmission distance, this protocol will enlarge the possible range for unconditionally secure key generation compared to the BB84 protocol [45].

## 5. Conclusions

In summary, we have demonstrated stable and backscattering-free operation of a PLC-based one-way QKD system with no active compensation. The possible key distribution distances are limited completely by the photon detector noise. Our experimental results clearly show the practical importance of one-way QKD systems for long distance key distribution. Our system is directly applicable to the BB84 protocol using a single photon source as well as the SARG04 protocol using WCP, which provide us with unconditionally secure keys.

**Acknowledgements**

The authors thank Satoshi Ishizaka for his technical assistance with the classical post-processing program. This work was supported by the National Institute of Information and Communication of Technology, Japan.

**Figure Captions**

**Fig. 1**. Schematic of PLC-based one-way QKD system. LD, 1.55-μm distributed feedback laser diode; AMZI, asymmetric Mach-Zehnder interferometer; ATT, attenuator; PM, phase modulator; and APD, avalanche photodiode. Dual-core fiber optic ribbon was used as an optical link between Alice and Bob. Bob's PM is polarization insensitive.

**Fig. 2**. Long-term stability of our PLC-based interferometer. Two AMZIs were initially set at temperatures at which maximum classical interference was observed and kept at those temperatures during the experiment. Temporal variation of the ratio of count rates in two output ports of the decoding AMZI is plotted. The inset shows the interference observed in count rates when the temperature of the encoding AMZI was changed.

**Fig. 3**. Polarization-insensitive operation of our PLC-based interferometer. Classical interference visibility is plotted against the temperature of the decoding AMZI. If we set the temperature to that at which maximum interference was observed, our interferometer is insensitive to polarization drift in the optical link.

**Fig. 4**. Experimental results of key distribution experiment. Top: key generation rate versus fiber length. Solid symbols represent the sifted key rate, and open symbols, the final key rate (see text for details). For all data, results with QE=5 and 10% are plotted. Broken lines are trend curves for the estimated final key rate using trend curves of QBERs. Bottom: measured QBERs versus fiber length. Broken lines are trend curves of QBER calculated using eq. (1).

**Fig. 5**.  Dark count probability per 1.2-ns gate window versus the quantum efficiency of the detector.

**Fig. 1**

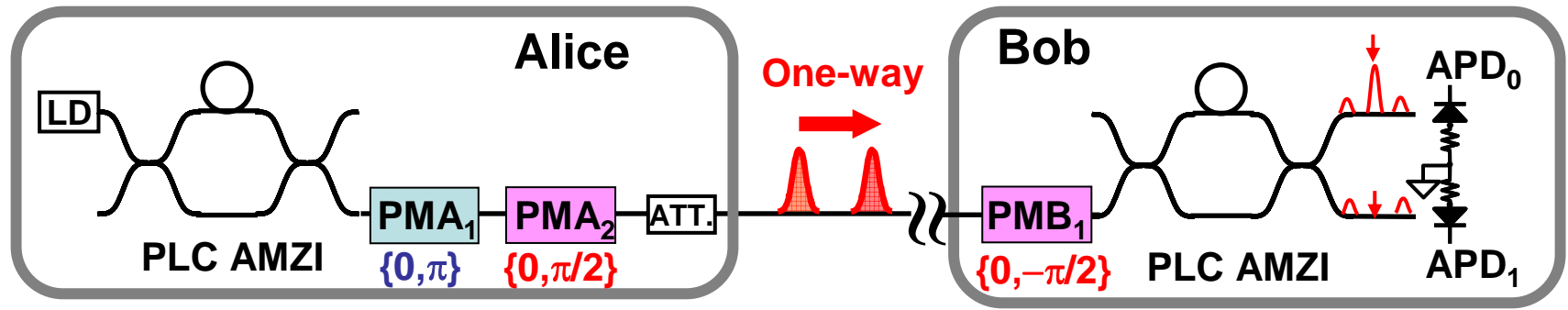

**Fig. 2**

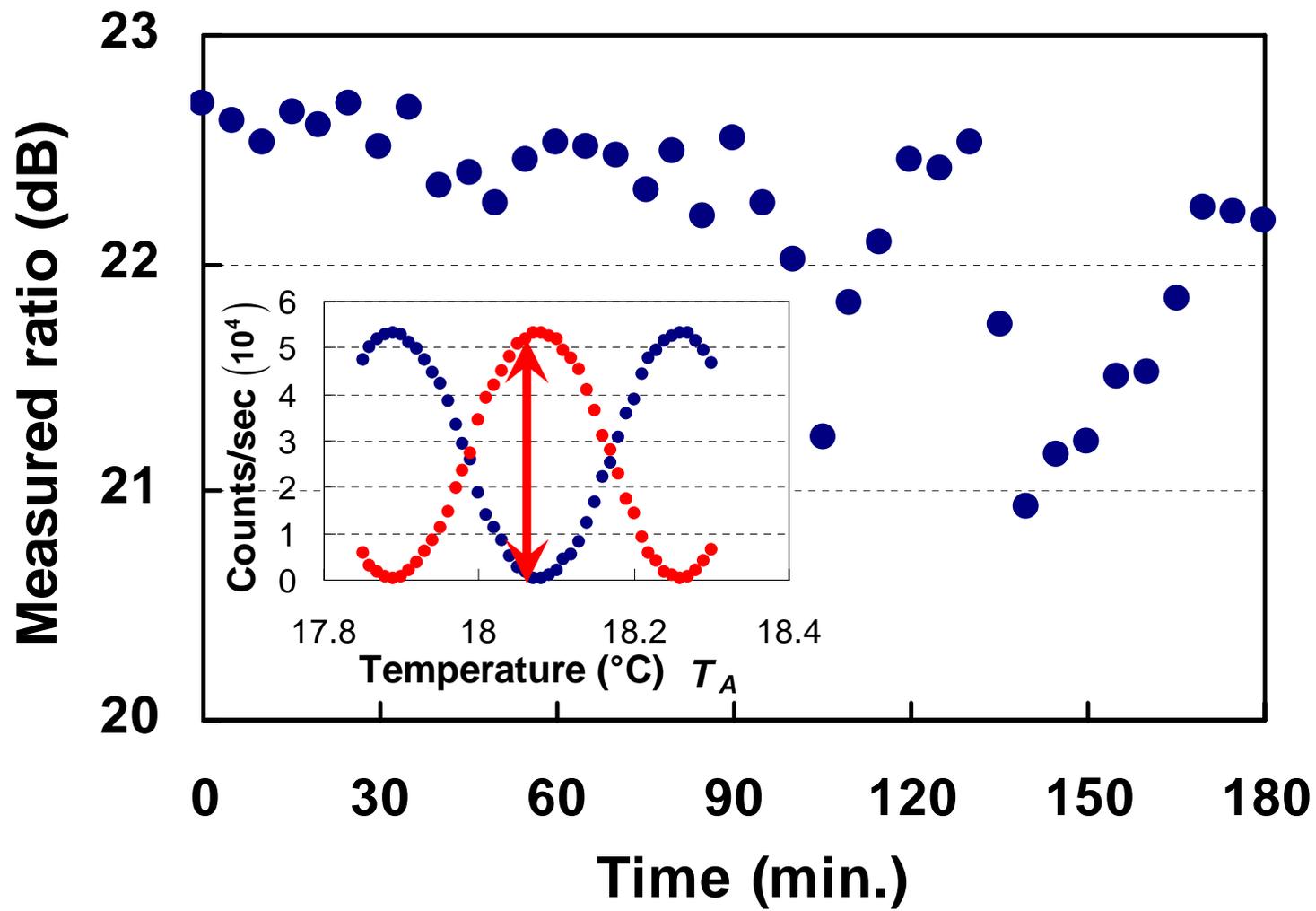

**Fig. 3**

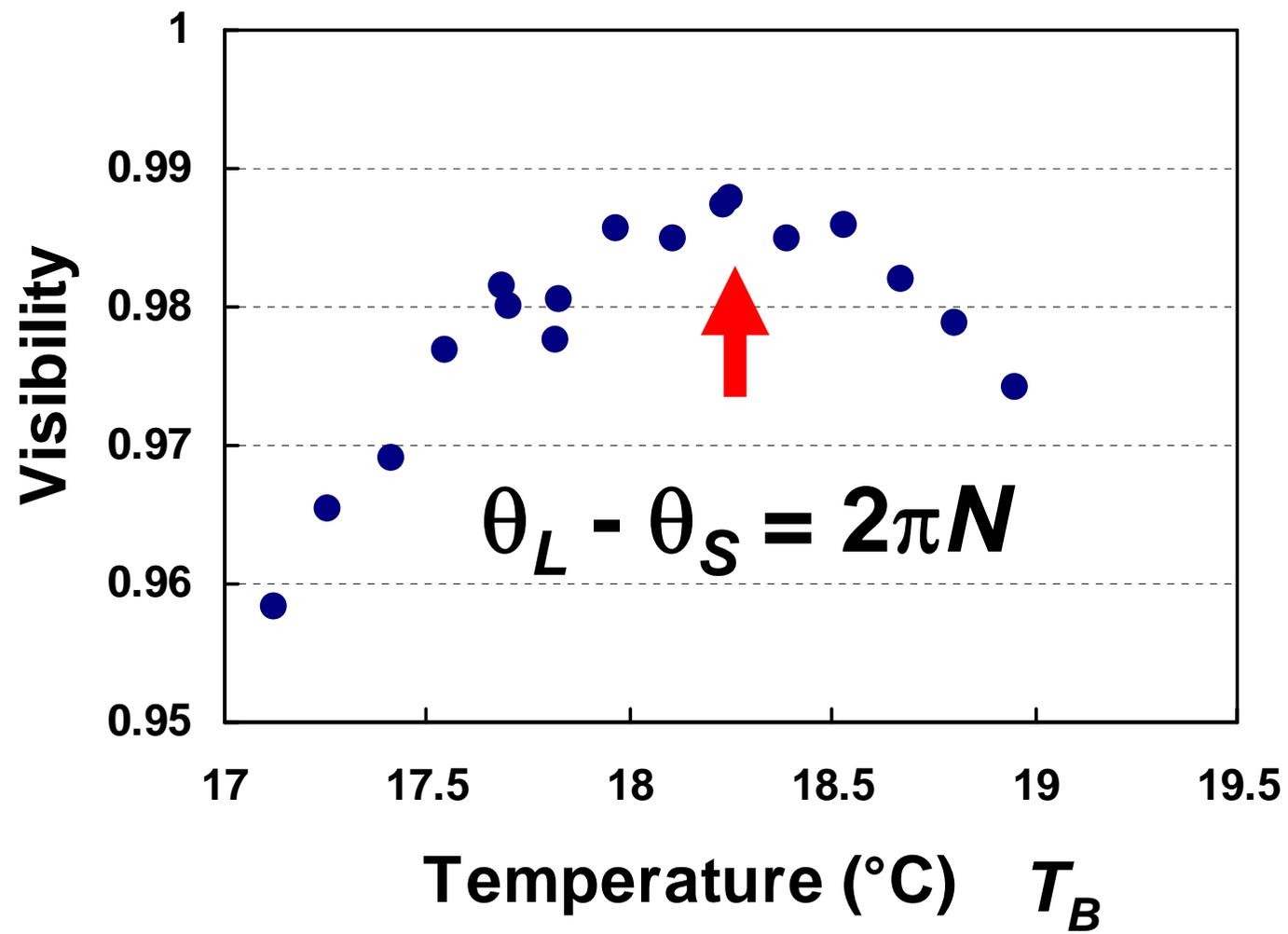

**Fig. 4**

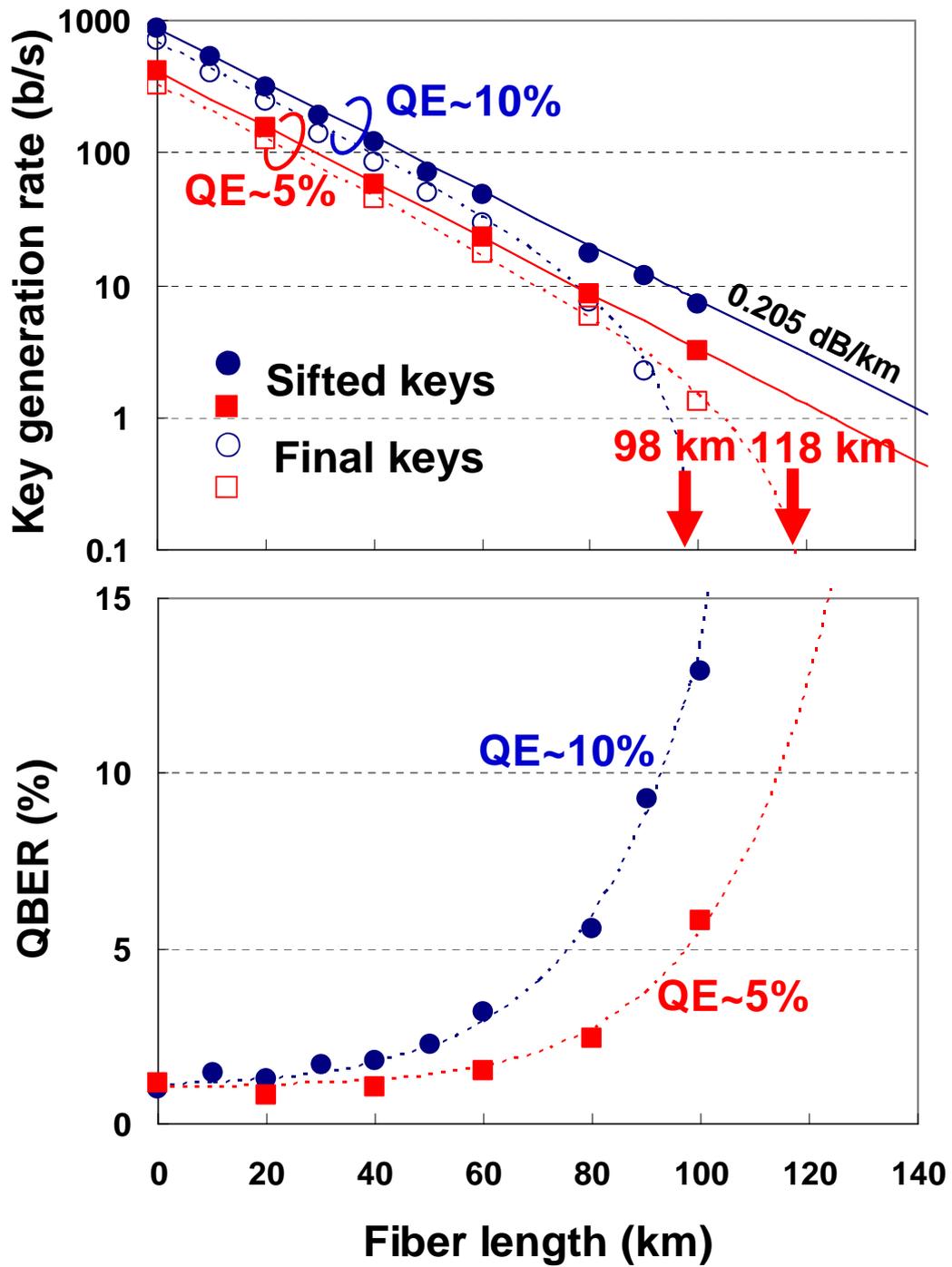

**Fig. 5**.

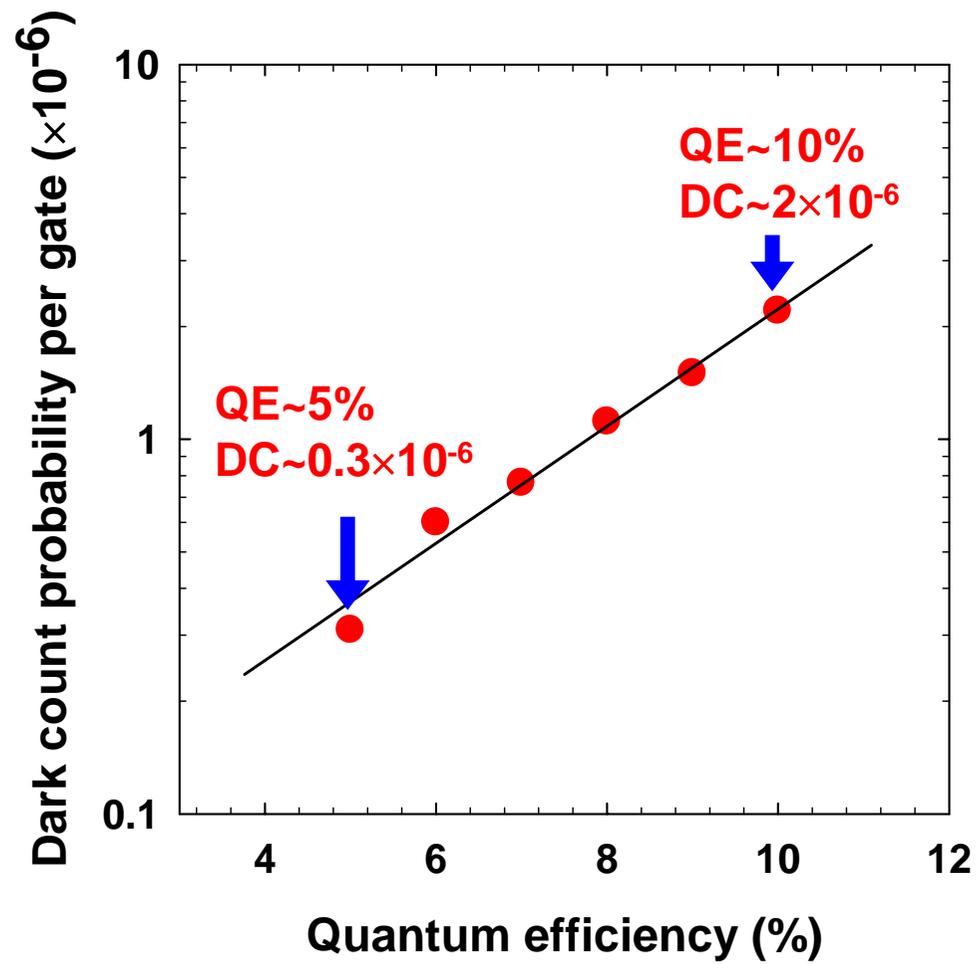